\documentclass{aa}
\usepackage{graphicx}
\usepackage{natbib}
\usepackage{amssymb}

\def\xmm{XMM-{\it Newton}}

\def\chandra{{\sc Chandra}\ }

\def\Cluster{2XMM\,J100451.6+411627}
\def\ClusterShort{XMM\,1004}

\def\xspec{{\sc Xspec}}
\def\mekal{{\sc MeKaL}}

\newcommand{\kpc}{ \ifmmode {\rm kpc} \else { kpc } \fi }
\newcommand{\keV}{ \ifmmode {\rm keV} \else { keV } \fi }

\newcommand{\Fig}[1]{Fig.\,\ref{#1}}
\newcommand{\Sec}[1]{Sec.\,\ref{#1}}
\newcommand{\Tab}[1]{Tab.\,\ref{#1}}

\def\ie{ {i.\,e,}\ }

\begin{document}

   \title{\Cluster\thanks{Based on observations obtained with 
                XMM-Newton, an ESA science mission with instruments and
                contributions directly funded by ESA Member States and NASA}: 
          A cool core cluster at $z=0.82$ }

   \author{     M. Hoeft\inst{1,2}
           \and G. Lamer\inst{2}
           \and J. Kohnert\inst{2}
           \and A. Schwope\inst{2} }

   \offprints{hoeft@tls-tautenburg.de}

   \institute{ $^1$ Th\"uringer Landessternwarte, Sternwarte 5, 07778 Tautenburg, Germany \\
               $^2$ Astrophysikalisches Institut Potsdam, An der Sternwarte 16, 14482 Potsdam, Germany\\
                }

   \date{Received; accepted}

   \abstract
       {}
       { 
       Gas cooling in the centre of massive galaxy clusters is believed
       to feed the most powerful active galactic nuclei in the Universe.
       How often clusters at high redshift show such cool cores has
       still to be explored by current and upcoming X-ray telescopes.
       }
       { 
       We correlated extended X-ray emissions from the second \xmm\
       source catalogue with SDSS data to particularly identify distant
       clusters. \Cluster\ is a candidate luminous enough to obtain its
       redshift and temperature from the X-ray spectrum. We also
       determine the surface luminosity profile and estimate the
       temperature in a few spherical bins. The analysis is complemented
       by a Subaru $g'r'i'$-image.
       }
       { 
       \Cluster\ has a redshift a redshift of $z=0.82\pm0.02$ and a
       temperature of $k_{\rm B}T = 4.2\pm0.4\:{\rm keV}$. A
       double-$\beta$ profile fit yields a highly concentrated surface
       brightness, $c_{\rm SB} = 0.32$, \ie\ the clusters hosts very
       likely a strong cool core. This is supported by the relaxed
       morphology of the cluster and an central temperature decrease.
       }
       {}


   \authorrunning{ M. Hoeft et al. }
   \titlerunning{  \Cluster\ }

   \maketitle

\section{Introduction}
\label{sec-intro}

   The space between galaxies in clusters of galaxies is filled with the
   hot thin intra-cluster medium (ICM). It is heated by merger shocks
   and adiabatic compression and cools by optically thin thermal plasma emission.
   Cooling times in the centre of clusters may be shorter than the
   Hubble time, and as a result some clusters form a cool core. Temperature
   profiles of local galaxy clusters allow us to distinguish clearly
   between cool core and non-cool core clusters \citep{leccardi:08,
   santos:08,pratt:07}.
   \\

   The abundance of cool cores at high redshift may shed light on their
   formation and on the interaction between the central active galactic
   nuclei (AGN) and the ICM. The number of cool core clusters at high
   redshift is still debated. \citet{vikhlinin:07} argued that galaxy
   clusters with redshift $z>0.5$ show a less steep profile at a radius
   of $0.05 \, r_{500}$ and concluded by comparison with the profiles of
   low redshift clusters cluster that cool cores are rare at high
   redshift. However, \citet{poole:08} found by numerical simulations
   that cool cores generally survive cluster mergers. Hence, they may
   also exists in an epoch of frequent mergers. Recently,
   \citet{santos:08} found by evaluating the concentration of the
   surface brightness for  sample of 15 distant galaxy clusters observed
   with \chandra\ that about 60\,\% of the high-$z$ clusters show at
   least a mild cool core.
   \\

   In this paper we report the discovery of a galaxy cluster at $z=0.82$
   in the second \xmm\ source catalogue (2XMM) that shows a multitude of
   signs for a strong cool core. In \Sec{sec-xray} we describe the X-ray
   properties of \Cluster, including the concentration of the
   surface brightness. In \Sec{sec-subaru} we describe the observation
   of the cluster region by the Subaru telescope.
   \\

   The concordance cosmological parameters $\Omega_{\rm M} = 0.27$,
   $\Omega_\Lambda = 0.73$, and $H_0 = 71 \: \rm km \, s^{-1} \,
   Mpc^{-1}$ are used throughout this paper. The corresponding linear
   scale at $z=0.82$ is $7.6 \, \rm kpc / ''$. All errors indicate the
   $1\sigma$ uncertainty level, except stated otherwise.

\section{X-ray data}
\label{sec-xray}

\begin{figure}
   \begin{center}
   \includegraphics[width=0.47\textwidth,angle=0]{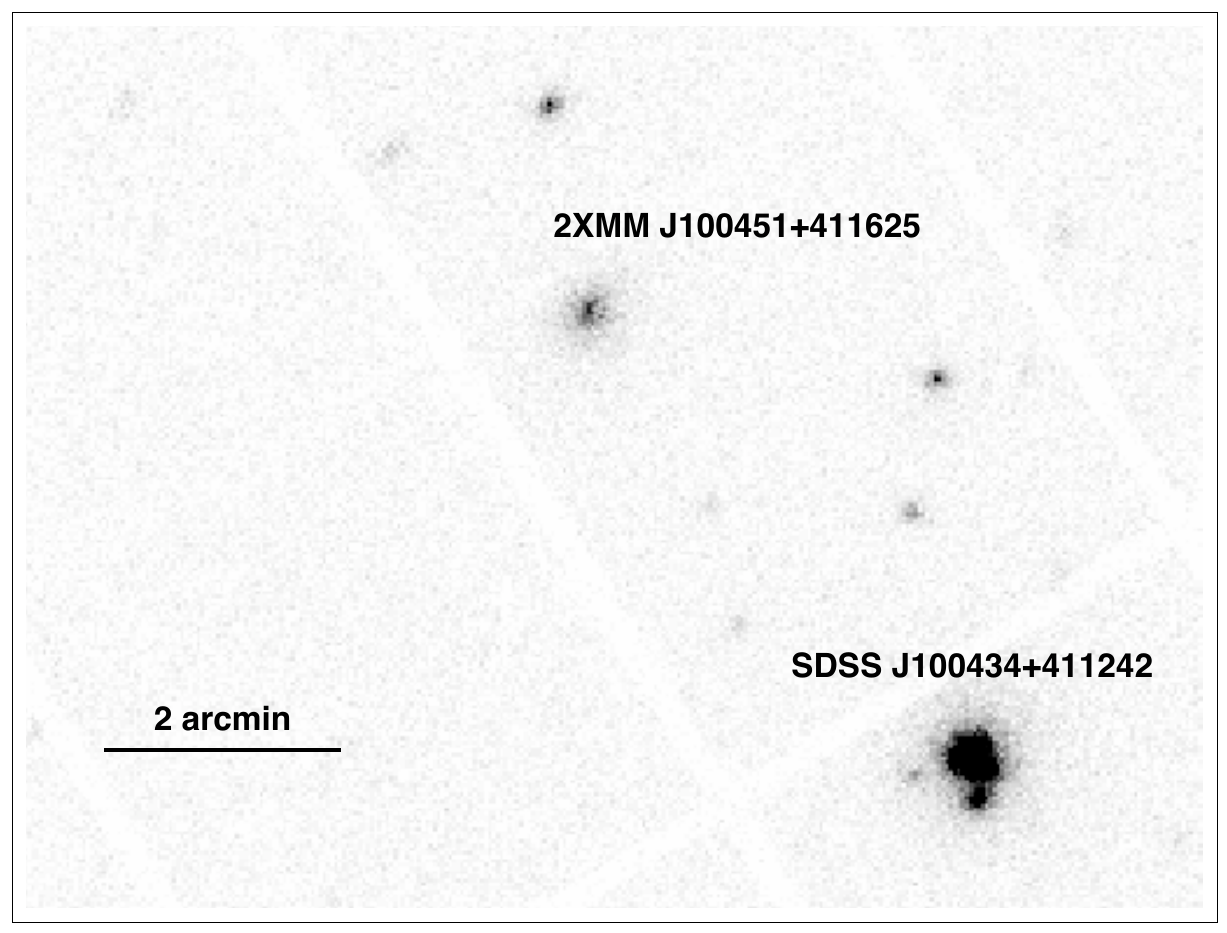}
   \caption{
      EPIC\,PN and MOS image of the observation 0207130201 
      in the $0.2-4.5\:{\rm keV}$ energy band.
      } 
   \label{fig-pn-mos-image} 
   \end{center}
   \end{figure}

   The galaxy cluster \Cluster\ (abbreviated as \ClusterShort\ in the
   following) was serendipitously detected in XMM\,EPIC observations of
   the lensed quasar SDSS\,J1004+4112 \citep[Observation ID
   0207130201,][]{lamer:06} and entered the 2XMM catalogue
   \citep{watson:08} as an extended X-ray source. By visual screening we
   found that \ClusterShort\ is clearly visible in the images of all
   three EPIC cameras as an extended source, see \Fig{fig-pn-mos-image}.
   The MOS images indicate that the X-ray luminosity of \ClusterShort\
   is very concentrated in the centre.
   \\

   The position of \ClusterShort\ is covered by the Sloan Digital Sky
   SurveyÊ(SDSS) and in the $i$-band image one galaxy is marginally
   detected close to theÊcentroid of the X-ray emission (SDSS\,
   100451.80+411626.6). Therefore the X-ray emission is most likey
   caused by a distant galaxy cluster.
   \\

   With 2500 source counts in the EPIC\,PN camera the source is bright
   enough to attempt a redshift determination based on the X-ray
   spectrum. We extract the overall cluster spectrum from a circle with
   a radius of $50''$ centred on the brightest pixel in the image of
   the EPIC\,PN camera. On the EPIC\,MOS cameras the chip gap is $23''$
   away from the cluster centre. However, we do not expect that the gap
   significantly affects the spectra and therefore include include all
   three cameras into the spectral analysis.
   \\

\begin{figure}
   \begin{center}
   \includegraphics[width=0.47\textwidth,angle=0]{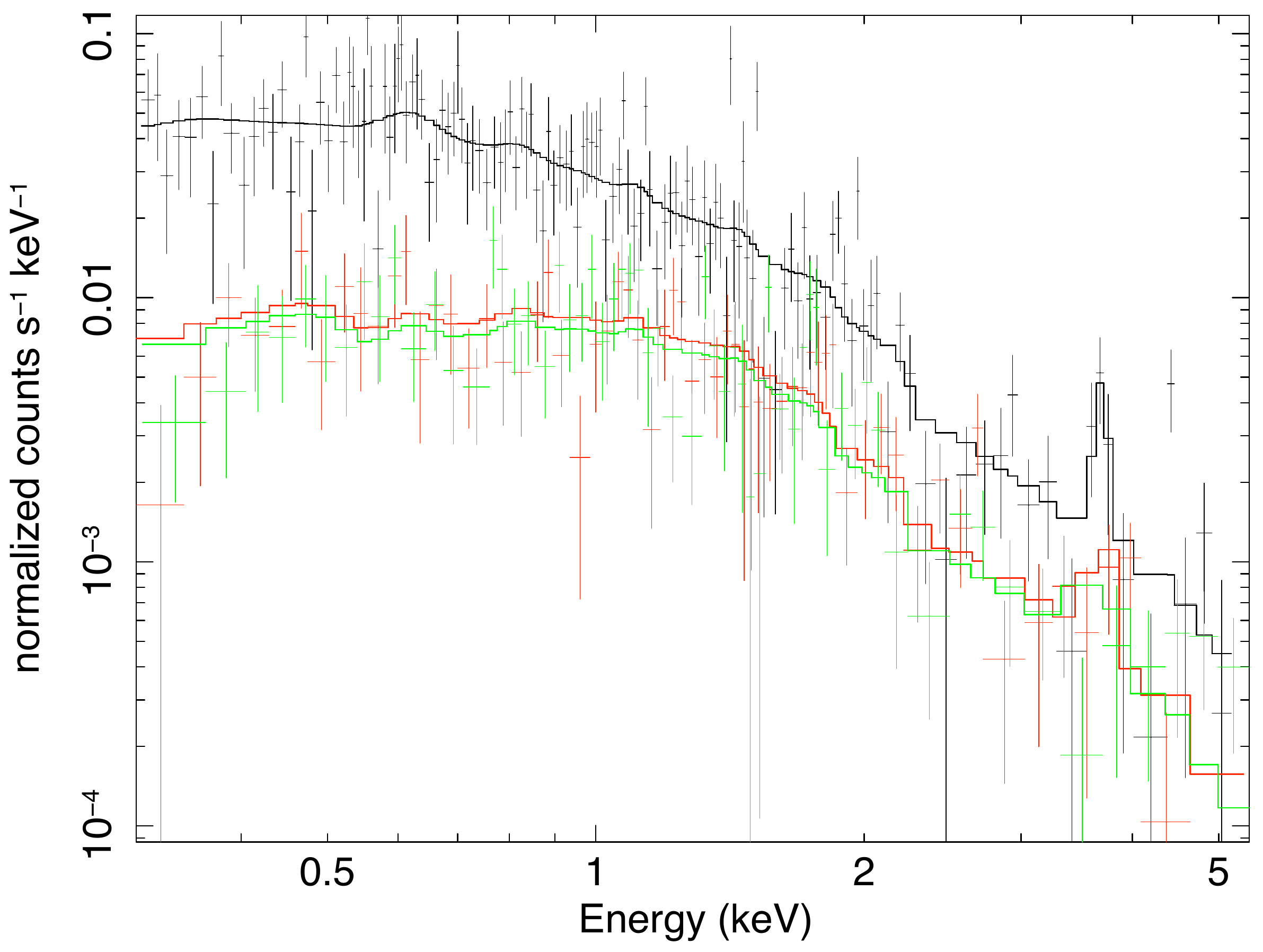}
   \caption{ EPIC\,MOS (lower spectrum red and green) and the PN (upper spectrum) data with
      best fitting \mekal\ model.
      } 
   \label{fig-spectrum} 
   \end{center}
   \end{figure}

\begin{figure}
   \begin{center}
   \includegraphics[width=0.47\textwidth,angle=0]{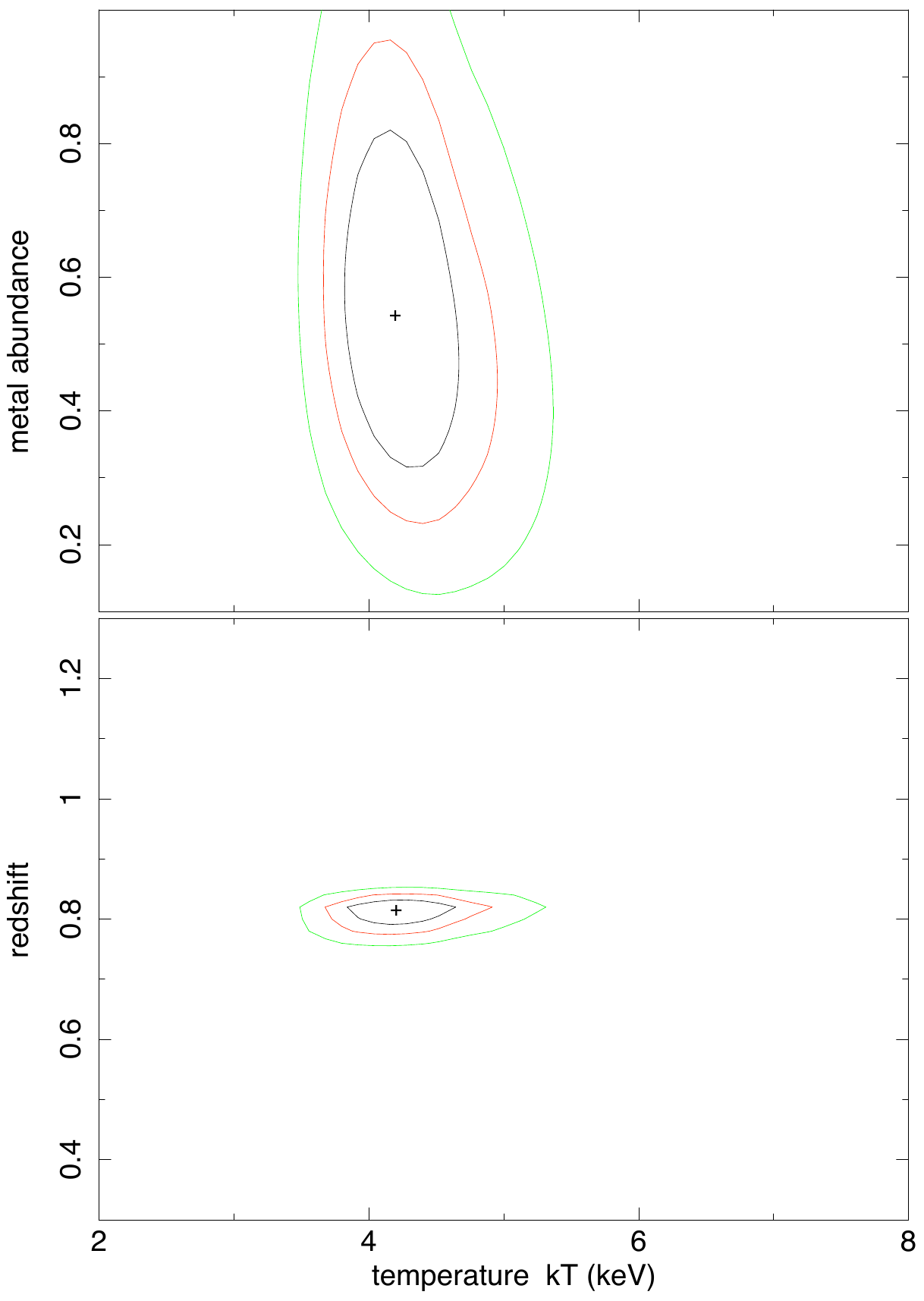}
   \caption{
       {\em Left}: 
       Confidence contours for a fit to the X-ray data using two free
       parameters, namely redshift and  temperature. Starting from the
       innermost contour, the levels indicate 68\%, 90\%, and 99\%
       confidence.
       {\em Right:} 
       Same confidence contours for metal abundance and temperature.
      } 
   \label{fig-contours} 
   \end{center}
   \end{figure}

   We use \xspec\ version 12.0 to fit a \mekal\ plasma model and to
   subtract the background spectrum. Fits are carried out using the
   Cash-statistics, which is preferable for low number counts per
   spectral bin. The galactic neutral hydrogen column density in the
   direction of \ClusterShort\ is $N_{\rm H} = 1.2 \times 10^{20}\, \rm
   cm^{-2}$, as obtained from radio data\footnote{\tt
   heasarc.gsfc.nasa.gov/cgi-bin/Tools/w3nh/w3nh.pl}. The Fe\,K-line is
   clearly visible at $k_{\rm B}T = 3.5 \, \keV$, see
   \Fig{fig-spectrum}. Hence, the redshift of \ClusterShort\ is
   unambiguously determined by the X-ray data. The best-fit redshift,
   temperature and metallicity are $z = 0.82\pm 0.02$, $k_{\rm B}T = 4.2
   \pm 0.4\,\keV$, and  $Z/Z_\odot = 0.5 \pm 0.3$, see
   \Fig{fig-contours}. 
   \\

\begin{figure}
   \begin{center}
   \includegraphics[width=0.47\textwidth,angle=0]{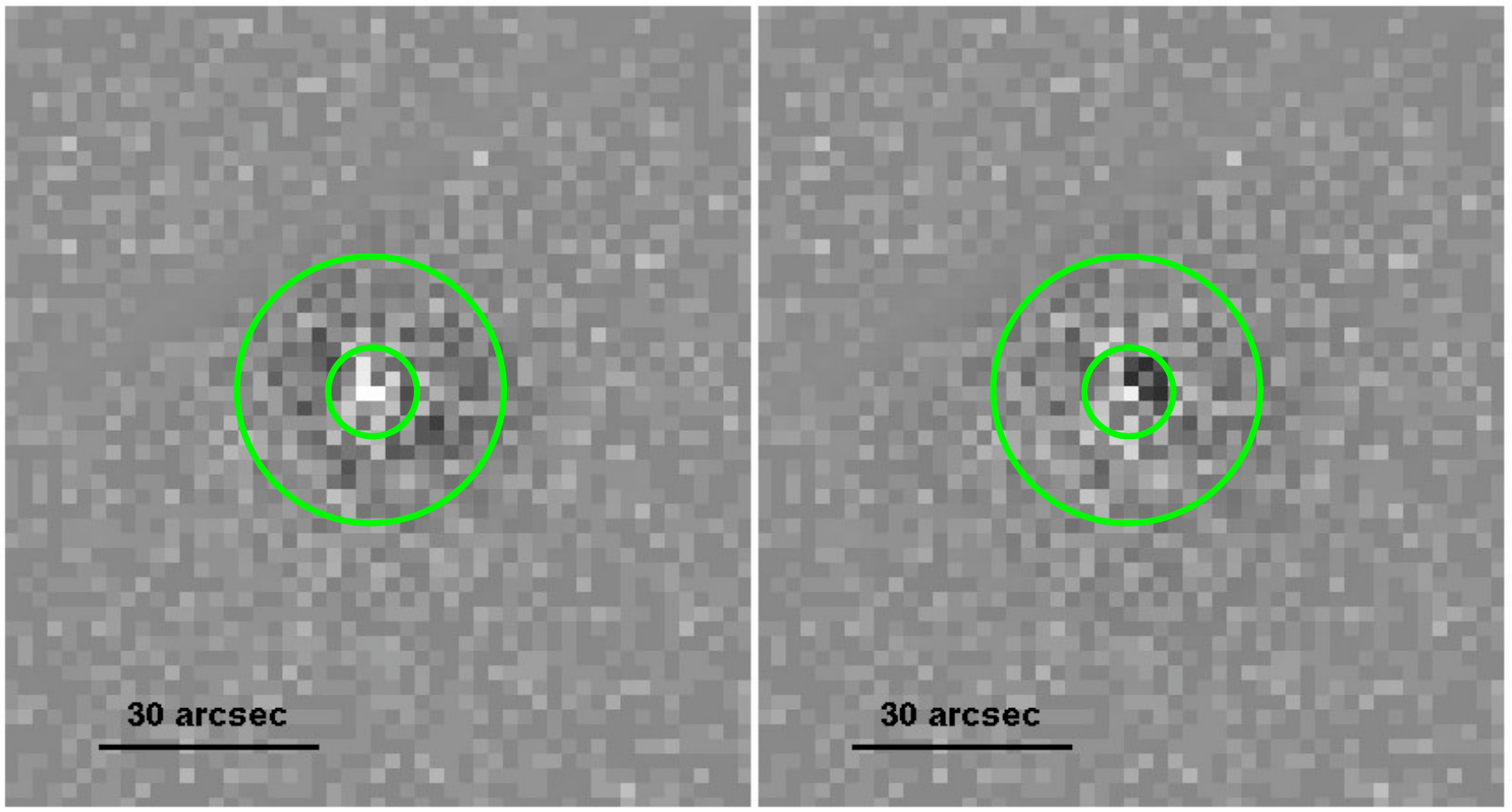}
   \caption{
      Residual images after fitting the combined EPIC MOS data with 
ÊÊÊ   a single-$\beta$ model ({\it left}) and a double-$\beta$ model ({\it right}) 
ÊÊÊ   to the combined EPIC MOS data  For the single-$\beta$
      we find an excess emission in the inner circle and a moderately
      deficient emission in the ring around the inner circle. For the
      double-$\beta$ the is no systematic trend 
      (only the absolute noise increases towards the centre).
      } 
   \label{fig-diff-images} 
    \end{center}
   \end{figure}

   Our visual inspection indicated that the X-ray luminosity is 
   concentrated to the cluster centre.
   To quantify the concentration of the surface brightness
   distribution of \ClusterShort\ we determined the surface brightness
   profile. Aiming to deconvolve the intrinsic cluster
   profile and instrumental point spread function (PSF), we
   performed a model fitting to the X-ray images. We have used a
   modifified version of the XMM-SAS task {\tt emldetect} to fit
   two-dimensional models of the X-ray brightness distribution to the
   combined EPIC\,MOS1 and MOS2 (0.2-4.5 keV) image of \ClusterShort. {\tt
   Emldetect} allows fitting $\beta$-models of the form 
   \[ I(x,y) \propto \left(1+\frac{(x-x_0)^2+(y-y_0)^2}{r_{\mathrm c}^2}\right)^{-3\beta+1/2} \]
   where $x$ and $y$ give the coordinates in the image plane and $\beta$
   is fixed at the canonical value $2/3$ for the surface brightness
   distribution of galaxy clusters. This model is convolved with
   the instrumental PSF and multiplied with the exposure map of the
   cameras in each fitting loop. Fitting the luminosity by a
   single-$\beta$ model resulted in a systematic discrepancy between the
   surface luminosity and the fit, see \Fig{fig-diff-images}. We therefore
   used a double-$\beta$ fit. A modification of {\tt emldetect} was
   necessary to allow fitting more than one extended source at then same
   source position, more precisely to fit a double-$\beta$ profile with two core
   radii. The model fitting was performed within a radius of $40''$
   around the centroid of the cluster X-ray emission. 
   \\

   \begin{table}[b]
   \begin{center}
   \begin{tabular}{ll}
   \hline
   \rule{1cm}{0mm}\\[-1.2ex]
   ObsID                                                       &  0207130201            \\[.4ex]
   PN paramaters in 2XMM \\
   \rule{1.7mm}{0mm}  ONTIME [s]                               &  41566                 \\
   \rule{1.7mm}{0mm}  OFFAX  [arc\,min]                          &  5.6                   \\
   \rule{1.7mm}{0mm}  CR .2-12 keV [s$^{-1}$]                  &  Ê$0.086 \pm 0.002$    \\[.4ex]
   PN flux in 2XMM\\
   \rule{1.7mm}{0mm}   .2-12 keV [$10^{-13}{\rm erg/cm^2 s}$]  &  $1.96 \pm 0.13 $ Ê    \\
   \rule{1.7mm}{0mm}   .5-2 keV [${10^{-13}\rm erg/cm^2 s}$]   &  Ê$0.81 \pm 0.02 $     \\
   \rule{1.7mm}{0mm}  COUNTS                                   &  $2448 \pm Ê62$ Ê      \\ [.4ex]
   double-$\beta$ model \\
   \rule{1.7mm}{0mm}   $r_{\rm c1}$  [arc\,sec]                &  $20.0 \pm 1.0 $       \\
   \rule{1.7mm}{0mm}   $\int {\rm d}x \, {\rm d} y \: I_1$     &  $0.0138 $             \\
   \rule{1.7mm}{0mm}   $r_{\rm c1}$  [arc\,sec]                &  $2.8 \pm 0.4 $        \\
   \rule{1.7mm}{0mm}   $\int {\rm d}x \, {\rm d} y \: I_2$     &  $0.0107 $             \\[.8ex]
   \hline
  \end{tabular}
  \caption{ 
    \ClusterShort: 2XMM catalogue source parameters and results from the double-$\beta$ 
    profile fitting using the modified XMM-SAS task {\tt emldetect}.
    }
  \label{tab-2xmm}
   \end{center}
  \end{table}

   When the model is constrained to a single radius, the best fitting
   model has a core radius of $r_{\rm c} = 8.2''$. A better fit is is
   achieved with a double-$\beta$ profile and core radii of $r_1 = 2.8'' $
   and $r_2=20''$, where the source with the smaller core radius
   contributes $44\,\%$ to the X-ray flux. These radii are consistent
   with those of nearby cool core clusters, see \citep{santos:08}.
   Unfortunately, the angular resolution of \xmm\ does not allow to
   exclude a point source as origin for the central luminosity excess.
   Hence, it might alternatively be caused by a luminous AGN, However,
   the Subaru images show no signs for an AGN in the brightest central cluster
   galaxy, see \Sec{sec-subaru}. 
   \\

   The background subtracted bolometric X-ray flux as obtained from the
   \mekal\ fit to all three cameras amounts to $F_{\rm bol} = 1.4 \times
   10^{-13} \: \rm erg \, s^{-1} \, cm^{-2}$ within the aperture of
   $50''$. Applying the concordance cosmological model allows us to
   calculate the intrinsic luminosity of the cluster. The used aperture
   corresponds to a radius of $380 \, {\rm kpc}$. In contrast, the
   commonly adopted cluster boundary, $r_{500}$, amounts to  $680 \pm 30
   \, {\rm kpc}$, as obtained from the temperature-radius relation given
   by \citet{ohara:07}. The error was estimated by the uncertainty
   in the temperature determination of \ClusterShort. From the
   double-$\beta$ model fit, see \Tab{tab-2xmm}, we derive
   $L(<r_{500})/L(50'') = 1.13$. As a result, we find for the
   intrinsic bolometric luminosity of \ClusterShort\ $L_{\rm
   bol}(<r_{500}) = 5.3 \times 10^{44} \: \rm erg \, s^{-1} $.
   \\

   According to the $L$-$T$ relation of local galaxy clusters, as
   derived by \citet{markevitch:98}, the temperature of \ClusterShort\
   corresponds to a luminosity of $2.4 \times 10^{44} \: \rm erg \,
   s^{-1} $. Thus, \ClusterShort\ is more than twice as luminous as expected
   from the low-redshift sample. The redshift evolution of the
   $L$-$T$ relation is still under debate. \citet{kotov:05} found that the
   luminosity scales with $(1+z)^{1.8}$, while \citet{ohara:07} report
   only a moderate evolution. \citet{pacaud:07} argued that the strong
   evolution found in earlier works is partially spurious due to
   selection effects. Taking these into account they found that the
   luminosity evolves in agreement with predictions based on
   self-similar cluster evolution. It predicts for $z=0.82$ a luminosity
   enhancement factor of 1.4, assuming the concordance cosmology model.
   Hence, scaling the luminosity of \ClusterShort\ according to
   self-similar cluster evolution keeps it over-luminous by a factor of
   1.5 compared to a local cluster sample. For local clusters it is
   known that the luminosity may be enhanced by a cool core,
   what may be also the case for \ClusterShort.
   \\

   \citet{santos:08} introduced the surface brightness concentration,
   $c_{\rm SB}$, as a measure for the presence of a cool cores. The
   concentration gives the ratio of the luminosities within 40 and  $400
   \, {\rm kpc}$. For \ClusterShort\ $40 \, {\rm kpc}$ corresponds to
   $5.3''$. We cannot determine the concentration from the cluster image
   directly, since the EPIC\,MOS cameras smooth the image substantially.
   More precisely, the FWHM of the PSF is $9''$ at the off-axis angle of \ClusterShort. 
   Therefore, we use the results of the double-$\beta$ model fit
   which provides a the deconvolved surface brightness profile. With this the
   surface brightness concentration of \ClusterShort\ amounts to $c_{\rm
   SB} = 0.32$. This is a higher concentration than \citet{santos:08}
   derived for any cluster in their sample including the local clusters.
   This clearly marks \ClusterShort\ as a candidate for a strong cool core cluster
   at high redshift.
   \\

\begin{figure}
   \begin{center}
   \includegraphics[width=0.45\textwidth,angle=0]{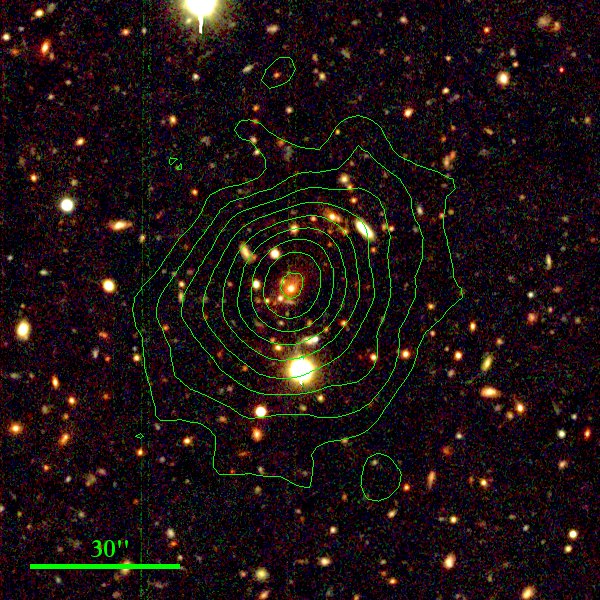}
   \caption{
       Subaru $g'r'i'$-colour-composite image with EPIC PN X-ray contours of
       \ClusterShort\ overlaid. 
       } 
   \label{fig-subaru} 
   \end{center}
   \end{figure}

\begin{figure}
   \begin{center}
   \includegraphics[width=0.47\textwidth,angle=0]{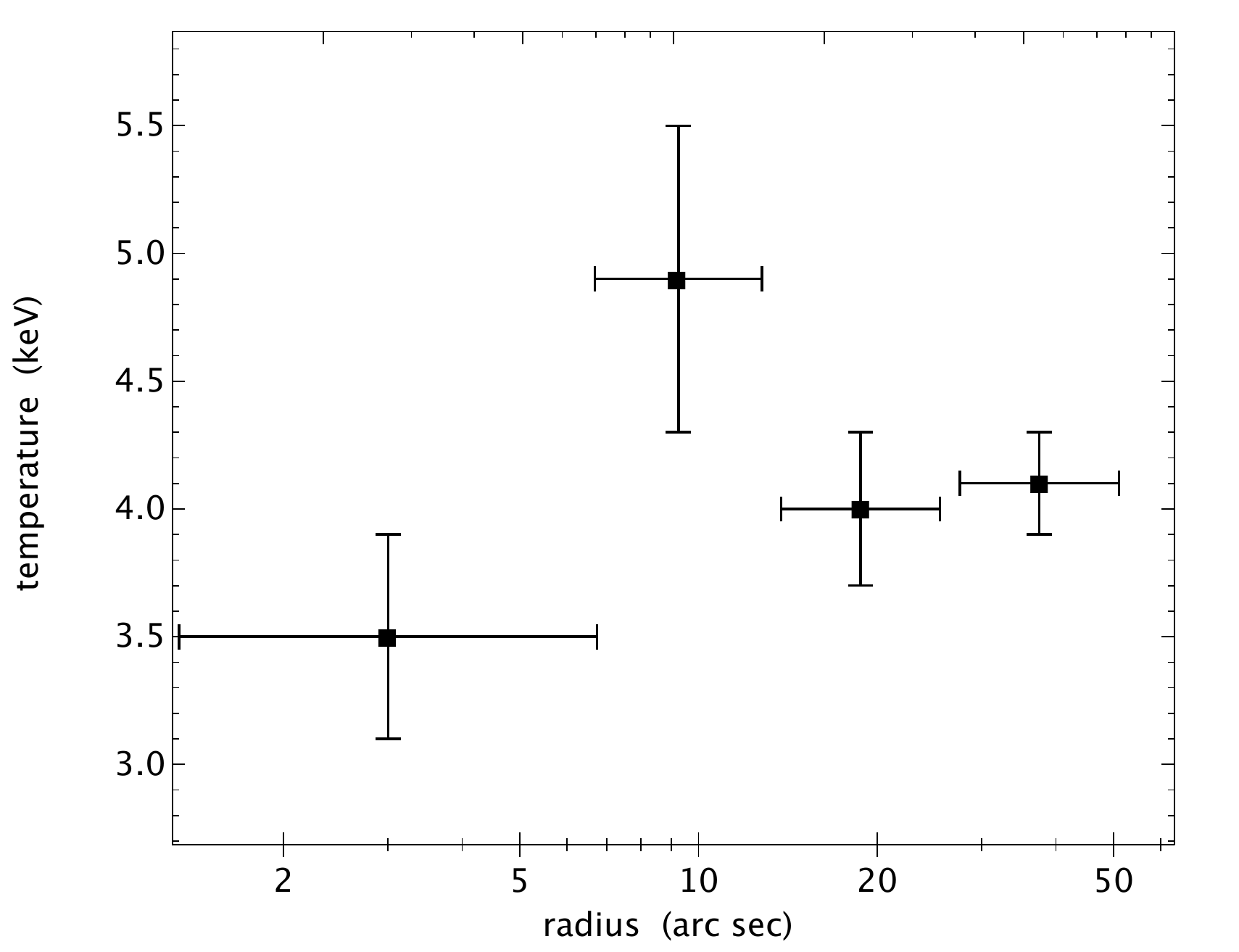}
   \caption{
      Temperature in four radial bin, using the PN and the MOS cameras.
      } 
   \label{fig-temperature-profile} 
   \end{center}
   \end{figure}

   Radial temperature profiles are routinely derived for local clusters
   \citep{pratt:07}. However, a large photon number is needed in order
   to extract the temperature from the spectrum computed for different
   regions. This may get even more difficult in merging clusters, which
   may show a very complex temperature distribution. Cool core clusters
   in contrast are generally relaxed systems, hence a spherically symmetric
   morphology is expected and is indeed found for
   \ClusterShort, see \Fig{fig-subaru}. Therefore we subdivide the area
   of \ClusterShort\ into four radial bins, enclosing in each bin a
   similar number of source photons. In each bin we fit the spectrum in
   a photon energy range from 0.3 to $7\:{\rm \keV}$, strictly excluding all
   time intervals with a flare of energetic particles, noticeable in the
   energy band above $10\:{\rm keV}$. We assume for all bins a common
   redshift and metallicity as derived for the entire cluster.
   \Fig{fig-temperature-profile}  shows the resulting temperatures
   in the four bins. In the inner region the temperature decreases by about 30\,\%,
   as it is known for low  redshift cool core clusters. \ClusterShort\
   is therefore the most distant cluster with a cool core verifiable by
   the temperature profile to date.
   \\

   The mass of a cluster can be estimated by assuming hydrostatical
   equilibrium and spherical symmetry and the average cluster temperature
   \[
     M_{500}
     \sim
     \frac{3 \beta}{G} \,
     \frac{ k_{\rm B} T \, r_{500} }  { \mu m_{\rm p} }
     .
   \]
   With $\beta = 0.66$ and $\mu=0.6$ we obtain for the cluster mass
   $M_{500} = (2.0 \pm 0.3) \times 10^{14} \: M_\odot$. The
   mass-temperature relation of \citet{vikhlinin:06} predicts $M_{500} =
   (1.4 \pm 0.2) \times 10^{14} \: M_\odot$, which is within $3\sigma$ of
   our derived value.
   \\

\section{Subaru images}
\label{sec-subaru}

   The field has also been observed with the Subaru Suprime-Cam imager
   in the ÊÊ$g'$, $'$, $i'$, and  $z'$-band, targeted on the lens system \object{SDSS
   J100434+411242}.  The respective exposure times were 810\,s, Ê1210\,s,
   1340\,s, and 180\,s. We retrieved the data from the {\sc Smoka} archive
   and used the SDFRED pipeline \citep{yagi:02} for the reduction of the
   $g'$, $r'$ and $i'$ images. We used point sources from the SDSS for the
   astrometric and photometric calibration of the Suprime-Cam data. The
   image quality of the reduced images is Ê${\rm FWHM} = 0.66''$ in the
   $g'$-band image, $0.78''$ in the $r'$-band and $0.73''$ in the $i'$-band. 
   \\
   
\begin{figure}[b]
   \begin{center}
   \includegraphics[width=0.47\textwidth,angle=0]{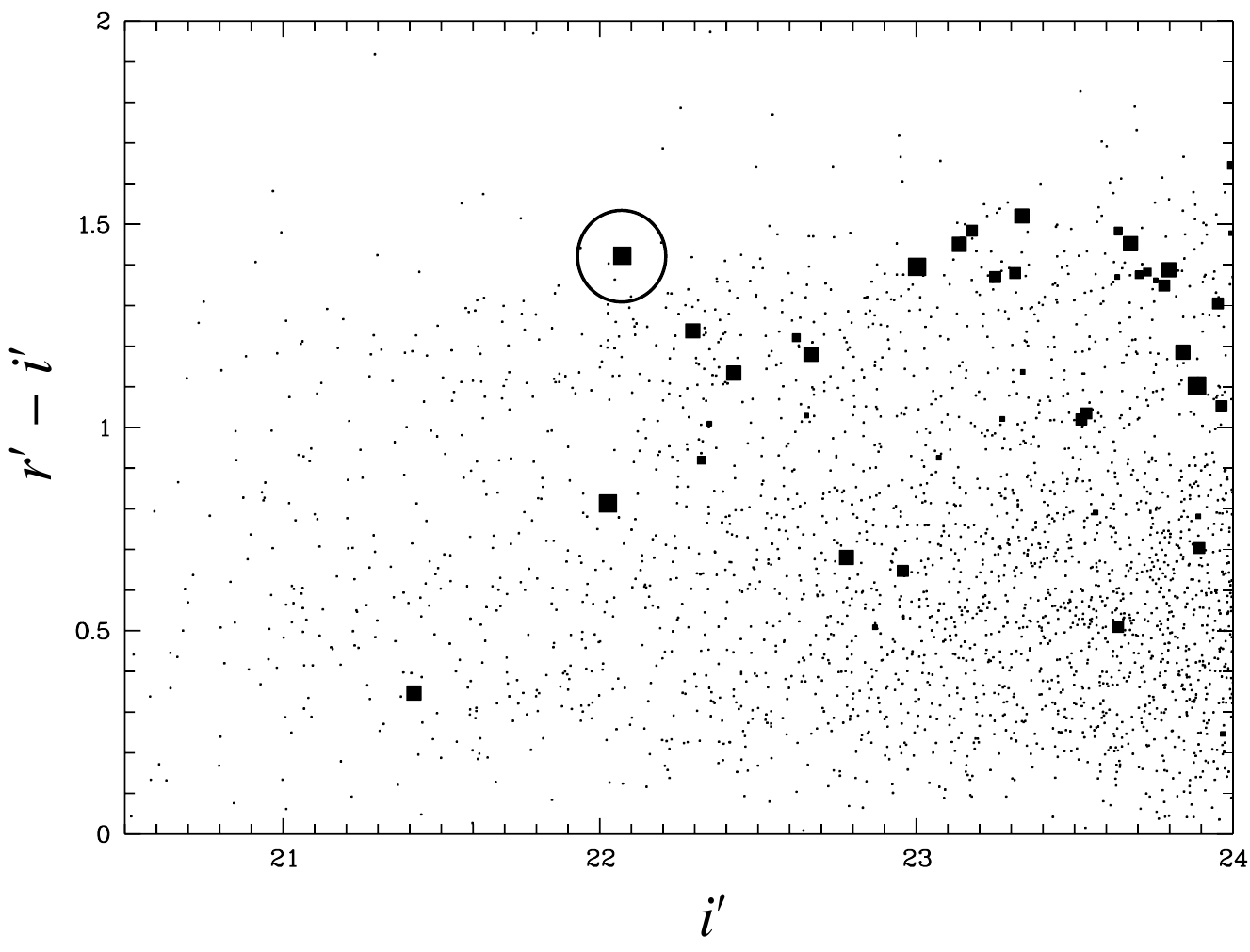}
   \caption{
      Colour-magnitude-diagramme of \ClusterShort. Larger symbols indicate
      lesser distance of the objects to the X-ray centre of the cluster.
      The BCG is encircled.
      } 
   \label{fig-cmd} 
   \end{center}
   \end{figure}

   The Subaru images reveal a clear overdensity of red galaxies within
   the X-ray contours. The brightest cluster galaxy (BCG) is located close to
   the centroid of the X-ray emission, see \Fig{fig-subaru}, indicating
   that \ClusterShort\ has a very relaxed morphology. The $r'-i'$ vs. $i'$
   colour-magnitude diagram, see \Fig{fig-cmd} Êof the galaxies close to
   the X-ray position shows a distinct red sequence with $r'-i '\sim 1.4$.
   The BCG is aligned with the red sequence, rendering a luminous AGN as source
   for the central X-ray luminosity excess unlikely.
   Ê

\section{Summary}
\label{sec-summary}

   Based on X-ray data we show that \ClusterShort\ is a luminous cluster
   at redshift  $z=0.82\pm0.02$ and with the help of a \mekal\ fit we
   determine its temperature, $k_{\rm B}T = 4.2\pm0.4\:{\rm keV}$. The
   surface luminosity of the cluster is exceptionally concentrated,
   $c_{\rm SB} = 0.32$, compared to other distant galaxy clusters
   \citep{santos:08}. This suggests that \ClusterShort\ hosts a strong
   cool core. A decreasing temperature in the centre and the extremely
   relaxed morphology of the cluster may serve as further pieces of
   evidence for a strong cool core cluster in the distant Universe.

\begin{acknowledgements}
    JK is supported by the DFG priority programme STP1177 (grant
    no.~SCHW563/23-1). MH acknowledge support by the Deutsches Zentrum
    f\"ur Luft- und Raumfahrt (DLR) under contract no.~FKZ 50 OX 0201.
    This work is based on observations obtained with XMM-Newton.
\end{acknowledgements}

   \bibliography{xmm1004_minimal}
   \bibliographystyle{apj}

\end{document}